\documentclass[review=false, sigchi]{acmart}

\usepackage{listings}
\usepackage{color}
\usepackage{pgfplots}
\usepackage{subfig}

\acmConference[IWOCL'20]{International Workshop on OpenCL}{April 27--29, 2020}{Munich, Germany}
\title[Architecture-Independent Memory Access Pattern Analysis]{Characterizing Optimizations to Memory Access Patterns using Architecture-Independent Program Features}

\author{Aditya Chilukuri}
\email{aditya.chilukuri@anu.edu.au}
\affiliation{%
	\institution{Australian National University}}

\author{Josh Milthorpe}
\email{josh.milthorpe@anu.edu.au}
\affiliation{%
	\institution{Australian National University}}

\author{Beau Johnston}
\email{beau.johnston@anu.edu.au}
\affiliation{%
	\institution{Australian National University}}

\acmYear{2020}
\acmMonth{6}
\acmPrice{15.00}


\definecolor{dkgreen}{rgb}{0,0.6,0}
\definecolor{gray}{rgb}{0.5,0.5,0.5}
\definecolor{mauve}{rgb}{0.58,0,0.82}

\usepackage{booktabs}
\usepackage{relsize}
\usepackage{multirow}
\usepackage{array}

\definecolor{tableheadcolor}{rgb}{0.8,0.8,1.0}
\definecolor{tablealtcolor}{rgb}{0.9,0.9,0.95}

\newcolumntype{B}{>{\global\let\currentrowstyle\relax}}
\newcolumntype{^}{>{\currentrowstyle}}

\newcolumntype{C}{>{\bfseries}}

\let\oldtabular\tabular
\let\endoldtabular\endtabular
\renewenvironment{tabular}{\sffamily\oldtabular}{\endoldtabular}

\begin{document}

	\begin{abstract}
	
	High-performance computing developers are faced with the challenge of optimizing the performance of OpenCL workloads on diverse architectures.
	The Architecture-Independent Workload Characterization (AIWC) tool is a plugin for the Oclgrind OpenCL simulator that gathers metrics of OpenCL programs that can be used to understand and predict program performance on an arbitrary given hardware architecture.
	However, AIWC metrics are not always easily interpreted and do not reflect some important memory access patterns affecting efficiency across architectures.
	We propose a new metric of parallel spatial locality -- the closeness of memory accesses simultaneously issued by OpenCL work-items (threads).
	We implement the parallel spatial locality metric in the AIWC framework, and analyse gathered results on matrix multiply and the Extended OpenDwarfs OpenCL benchmarks.
	The differences in the observed parallel spatial locality metric across implementations of matrix multiply reflect the optimizations performed.
	The new metric can be used to distinguish between the OpenDwarfs benchmarks based on the memory access patterns affecting their performance on various architectures.
	The improvements suggested to AIWC will help HPC developers better understand memory access patterns of complex codes and guide optimization of codes for arbitrary hardware targets.
	\end{abstract}

	\maketitle	
	
	\section{Introduction}
	
	High-performance computing (HPC) systems are increasingly heterogeneous.
	A single node on a modern supercomputer may combine traditional CPUs with one or more accelerators such as GPUs, Field Programmable Gate Arrays (FPGAs), or many integrated core devices (MICs).
	High bandwidth interconnects support tight integration between multiple devices of different types on a single compute node.
	
	The OpenCL programming language is designed to support modern HPC software engineers in writing code that executes on multiple hardware targets. 
	This gives HPC software developers greater flexibility by allowing codes aimed at a range of hardware targets to be written in a single programming language environment.
	
	Application codes differ in resource requirements, control structure and available parallelism. Similarly, compute devices differ in number and capabilities of execution units, processing model, and available resources.
	These heterogeneous computing environments present opportunities for software engineers and HPC integrators to design highly optimized systems with multiple kernels executing on hardware targets best suited for the diverse computational tasks performed by each kernel~\cite{spafford2010maestro}.
	However, this opportunity also presents a challenge in optimizing code to run on diverse architectures.
	The work reported here aims to help HPC developers understand and optimize memory access patterns to improve performance on heterogeneous systems.
	
	In the evolution of computer architectures over the last few decades, the exponential growth in computational capability has not been matched by proportional increases in memory speeds\cite{hennessycomparch}.
	Memory accesses pose larger bottlenecks to performance as application demand for main memory scales with the arithmetic capability of computer systems.
	To mitigate this latency, modern CPU designs have employed a wide range of cache technologies to reduce main memory accesses using the principle of \emph{spatial locality}: the observation that data that a program accesses close together in time tend to also be close together in memory.
	On the other hand, GPUs rely on hardware multithreading to hide memory latency, and their architectures favour ALU capability over sophisticated logic to manage a cache hierarchy and out-of-order execution. As a result, the performance of kernels on CPUs and GPUs alike is strongly dependent on memory access patterns intrinsic to the code. 
	

	Our aim in this paper is to develop a framework to guide HPC software engineers in hardware-dependent code optimization -- specifically by guiding the improvement of memory access patterns.
	We provide examples of manufacturer-recommended code optimizations to improve memory access patterns on the target architecture, and examine these using the architecture-independent workload characterization (AIWC)~\cite{beauaiwc} plugin for Oclgrind~\cite{price:15}.
	Our work highlights the benefits and challenges arising from an architecture-independent analysis of memory-based program characteristics.
	We propose two new metrics for AIWC to better characterize these memory-based optimizations.
	We measure the presented codes using the new metrics and demonstrate the metrics' effectiveness in capturing the essence of the optimizations performed.
	
	The structure of this paper is as follows.
	In section~\ref{motivating example} we use the example of matrix multiplication on a GPU to showcase the specific vendor-recommended optimizations our work aims to measure.
	In section~\ref{related work} we discuss how AIWC and its precursors profiled memory access patterns in an architecture-independent fashion and also consider relevant architecture-dependent approaches to memory access profiling.
	In section \ref{method} we consider non-parallel memory metrics collected by AIWC and evaluate the suitability of these metrics for capturing the impact of performance optimizations to various OpenCL codes.
	In section~\ref{discussion} we propose a new parallel spatial locality metric, implement it in the AIWC framework, and evaluate it over the matrix multiplication example from section~\ref{motivating example}.
	In section~\ref{results} we present the collected metric for selected benchmarks from the Extended OpenDwarfs benchmark suite to validate our methodology in this paper.
	Finally, in section~\ref{future work} we discuss further avenues to extend our work and conclude.
	
	\section{Motivating Example} \label{motivating example}
	
	We aim to capture the essence of vendor-recommended optimizations for target architectures in our metrics.
	We demonstrate the optimization strategies using an OpenCL kernel which multiplies square matrices of order $N$.
	We start with a simple unoptimized kernel, and then improve the memory access patterns of this kernel by incrementally performing the optimizations recommended in the CUDA Optimization Handbook \cite{cudaoptimisation} for NVIDIA GPUs.
	These incrementally optimized OpenCL codes are then analysed using AIWC to determine the accuracy in profiling favourable memory access patterns of the current framework and the proposed extensions to AIWC.
	
	\subsection{Simple Unoptimized Matrix Multiplication}
	
	The unoptimized matrix multiplication kernel presented in Appendix~\ref{simpleMultiply} is used as a baseline to validate the performance improvements measured from each following optimization.
	Each thread of the kernel updates the $(globalRow, globalCol)$ element of matrix $C$ by computing the dot product of the corresponding row of $A$ and column of $B$.
	
	\subsection{Using Shared Memory to Coalesce Global Memory Access to Matrix A} 
	
	We first notice that the number of global memory accesses to matrices A and B increases as $O(N^3)$ with respect to matrix size $N$.
	Global memory is typically located off-chip and accesses induce large delays.
	NVIDIA GPUs coalesce global memory loads and stores issued within thread-groups into as few DRAM transactions as possible.
	Multiple global memory loads and stores are coalesced into a single transaction when certain device-specific conditions are met. 
	On most NVIDIA GPUs, data accesses are coalesced when multiple requests are made for memory locations from the same cache line in global memory~\cite{cudamanual}.
	Appendix~\ref{coalescedAMultiply} contains the \texttt{coalescedA} kernel, which coalesces accesses to matrix $A$ by storing \emph{tiles}, or square blocks, of $A$'s values into shared memory (OpenCL \texttt{\_\_local} memory). 
	
	
	\subsection{Using Shared Memory to store Tiles of Matrix B} 
	
	The code is further optimized by improving the locality of reads from Matrix $B$.
	While in the previous kernel each thread reads only a single element of matrix $A$ from global memory, each thread reads a full column of matrix $B$.
	The repeated reads of elements of matrix $B$ can be shared between threads in a work group by reading tiles from matrix $B$ into shared memory.
	Appendix~\ref{coalescedABMultiply} contains the \texttt{coalescedAB} kernel which performs this optimization.
	
	\subsection{Optimizing Handling of Shared Memory} 
	
	NVIDIA shared memory is divided into multiple banks -- stored in independent memory modules -- to allow parallel memory access.
	Bank conflicts occur when shared memory in the same bank is accessed concurrently.
	The code in \texttt{coalescedAB} kernel is further optimized by implicitly transposing tiles of Matrix $A$ while loading from global memory. This improves memory bank utilization during reads to shared memory. The \texttt{coalescedABT} kernel demonstrates this optimization by modifying lines 20,21 of \texttt{coalescedAB} to:
	
	\begin{lstlisting}[firstnumber=20]
		ASub[localCol][localRow] = A[tiledRow];
		BSub[localRow][localCol] = B[tiledCol];
	\end{lstlisting}

	\subsection{Alignment of Memory Allocation}
	
\begin{figure}[h]
		\includegraphics[width=\linewidth]{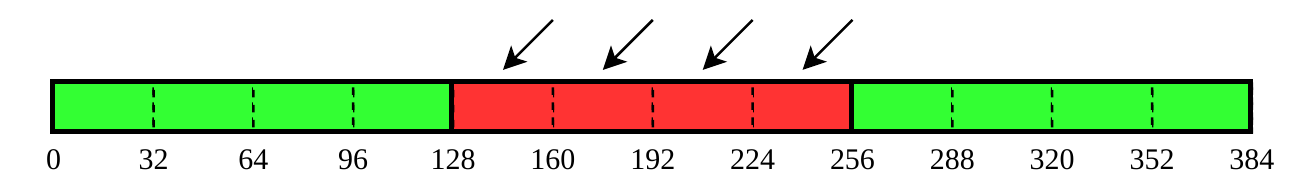}
		\caption{All threads access memory aligned to nearest cache line in parallel \cite{cudaoptimisation}}
		\label{fig: Simple memory alignment}
	\end{figure}

\begin{figure}[h]
		\includegraphics[width=\linewidth]{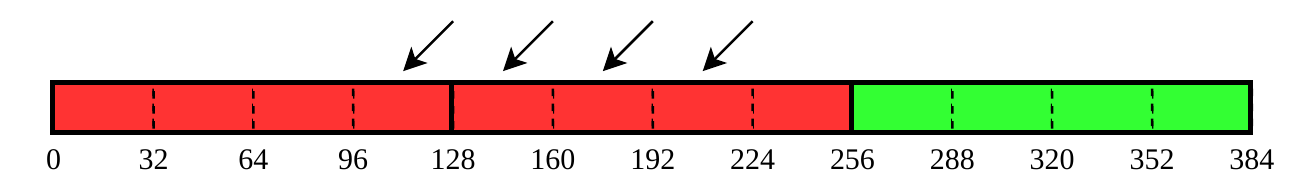}
		\caption{Unaligned sequential memory addresses fit in two cache lines \cite{cudaoptimisation}}
		\label{fig: Simple memory misalignment}
	\end{figure}

	Memory access alignment is important to best utilize all parts of GPU memory architecture. Global memory buffer alignment can allow threads to access blocks of global memory aligned to the nearest cache line. This enables coalescing of memory accesses. If buffers are misaligned, parallel memory requests may cross over cache lines, and may double the number of slow global memory accesses needed as demonstrated in figures \ref{fig: Simple memory alignment} and \ref{fig: Simple memory misalignment}.
	
	A similar principle applies when using shared memory. Bank conflicts may be reduced by aligning allocations of shared memory buffers. The \texttt{alignedABT} kernel improves the alignment of shared memory tiles in \texttt{coalescedABT}.
	An arbitrary large alignment value of 4096 is chosen as it is larger than the cache line size of modern hardware and attributed to the local array declarations in the existing code as shown below.
	
	\begin{lstlisting}
		__local float aTile[TILE_DIM][TILE_DIM] __attribute__ ((aligned(4096)));
		__local float bTile[TILE_DIM][TILE_DIM] __attribute__ ((aligned(4096)));
	\end{lstlisting}
	
	All the examples above are an incomplete listing of optimization strategies developers can apply to code targeted at NVIDIA GPUs.
	In section~\ref{method}, we will compare the collected AIWC metrics with the expected effect of each optimization to examine how effectively our methodology uncovers underlying bottlenecks and guides optimization efforts.
	
	\flushbottom
	\section{Related Work} \label{related work}
	
	Hoste and Eeckout \cite{hoste2007microarchitecture} show that while conventional architecture-dependent characteristics are useful in locating performance bottlenecks, they can hide the underlying, inherent program behaviour causing performance bottlenecks.
	A conceptual understanding of the performance characteristics of complex codes is necessary for the programmer to effectively optimize these codes.
	Architecture-dependent characteristics typically include instructions per cycle (IPC), cache and branch prediction miss rates, page faults and DRAM bus data transfer rates. These are typically collected using hardware performance counters available on most target architectures. These performance counters do not serve to guide optimization beyond highlighting potential bottlenecks \cite{hoste2007microarchitecture, ganesan2008performance}.
	Further, in many cases, architecture-dependent characteristics cannot be directly correlated to specific code patterns.
	For example, the causes of high cache miss rates in the execution of a program are complex and depend on microarchitecture specific features such as cache size, prefetch behaviour and cache placement policies.
	A HPC developer tasked with optimizing code for a given hardware target would benefit from architecture-independent metrics of the code that can be used to measure the effect of code modifications.
    These metrics would help to guide the developer in both finding and fixing performance bottlenecks.
	
	To address the limitations of conventional microarchitecture-dependent characteristics, Hoste and Eeckout~\cite{hoste2007microarchitecture} developed the Microarchitecture-Independent Workload Characterization tool (MICA).
	They observed that performance counter-based approaches to profiling codes often failed to find underlying program features that map to improved or worsened usage of performance-critical hardware features of the target architecture.
	The MICA framework is a holistic characterization tool, and thus collects features including instruction mix, instruction-level parallelism, register traffic, data stream strides and branch predictability.
	
	Of these metrics, data stream strides are of particular interest in memory access pattern profiling.
	MICA's stride length metric measures the distance between consecutive memory accesses in a single-threaded application.
	For CPU architectures running single-threaded applications, this metric correlates to the spatial locality of memory accesses -- a measure of how closely bunched are memory access in nearby times.
	This is directly correlated to cache reuse rates, critical to code performance on CPUs \cite{inteloptimisation}. 
	
	The MICA approach was tailored for single-threaded applications as the metrics collected rely heavily on Pin instrumentation \cite{luk2005pin}.
	As such, MICA is unsuited to analysing HPC workloads with heavy use of parallelism.
	The Workload ISA-Independent Characterization for Applications (WIICA)~\cite{shao2013isa} extends MICA to present a framework to analyse single-threaded programs independent of the instruction set architecture (ISA).

	Kim and Shrivastava \cite{kim2011cumapz} present CuMAPz, a CUDA memory access profiling tool to guide NVIDIA GPU optimizations.
	CuMAPz focuses on the problem of improving CUDA application performance using NVIDIA memory-based optimizations.
	CuMAPz analyses CUDA codes structurally and simulates code execution on the memory hierarchy of specific NVIDIA GPU models.
	
	During simulation of the target code, CuMAPz records all memory accesses in various buffers to simulate the global and shared memories on NVIDIA GPUs.
	Using this detailed simulation data, CuMAPz can estimate the performance-critical (i) shared memory data reuse profit, (ii) profit from coalesced access (iii) memory channel skew cost and (iv) bank conflict cost characteristics of the target code.
	
	The simulation environment used by CuMAPz and the attached analysis framework is highly specific to CUDA enabled GPUs.
	Replicating the CuMAPz framework for all target architectures is challenging.
	However, CuMAPz is an interesting simulator from the standpoint of this paper's work since it adheres to the core parts of NVIDIA GPUs' memory models in its analysis, while allowing the user to specify their GPU model specific hardware information.
	
	Our approach is an architecture-independent analysis of memory access patterns to provide metrics correlating to similar performance critical memory access optimizations as CuMAPz.
	We aim to further the state of the art by providing a framework to guide developers in optimizing OpenCL codes for any given target architecture.
	To the best our knowledge, none of the previous works presents a set of performance metrics that accurately characterize memory access patterns of parallel applications independent of the target architecture.
	
	The Architecture-Independent Workload Characterization (AIWC) tool \cite{beauaiwc} collects a set of instruction set architecture (ISA)-independent features based on those identified by Shao and Brooks \cite{shao2013isa}.
	AIWC runs as a plugin to the Oclgrind~\cite{price:15} framework, which simulates OpenCL kernel execution on an ideal device according to the OpenCL execution and memory models.
	AIWC collects metrics of kernel memory access, including simple counts such as the (i) total memory footprint, the total number of unique addresses accessed; and (ii) 90\% memory footprint, the number of unique addresses covering 90\% of memory accesses.
	While these metrics are architecture-independent, they are correlated to program performance on typical architectures.
	For example, a small ratio of 90\% memory footprint to total memory footprint indicates that a program accesses a small subset of memory addresses frequently, which is highly beneficial for performance in a cached memory hierarchy.
	
	AIWC also records the global memory address entropy (GMAE), a positive real number corresponding to the randomness of the memory access distribution of a program. To measure locality of memory accesses, AIWC collects the local memory address entropy (LMAE) of memory addresses accessed after dropping $n$ least significant bits of all memory addresses accessed by the program. To calculate this, AIWC collects a frequency distribution of all non-register memory accesses by all threads in the target kernel. Using the collected frequency distribution, AIWC calculates 10 separate local memory address entropy (LMAE) values according to increasing number of least significant bits (LSB) skipped using
	the explicit formula for the $n$-bits skipped LMAE:
	
	\begin{equation}
	LMAE_{n-bits} = \sum_{a \in A_n}p_a \log_2(p_a^{-1})
	\end{equation}
	
	\begin{itemize}
		\item $A_n$ is the set of all addresses accessed after skipping $n$ LSBs of each address.
		\item $p_a := \frac{\#access_a}{\#access_{total}}$ is the probability (calculated as relative frequency) at which each memory address is accessed.
	\end{itemize}
	
	LMAE measures the locality of memory accesses performed over the full execution of a program. A steeper drop in entropy with increasing number of bits may correlate to more localized memory accesses over the program's execution.
	
	\begin{figure*}[ht!]
		\centering
		\subfloat{
			\centering
			\begin{tikzpicture}
			\begin{axis}[symbolic x coords={simple,coalA,coalAB,coalABT,alignABT},
			xtick={simple,coalA,coalAB,coalABT,alignABT},
			xticklabel style={text height=2ex},
			ymin=0,
			ylabel={Execution time (ms)}, xlabel={Kernel},
			ylabel style={at={(axis description cs:0.1,.5)},anchor=south},
			legend entries={$N=80\ \ \ \ $,$N=256\ \ $},
			legend style={at={(0.6,0.45)},anchor=west}
			]
			
			\addplot coordinates {
				(simple,.031)
				(coalA,.019)
				(coalAB,.020)
				(coalABT,.018)
				(alignABT,.018)
			};
			\addplot coordinates {
				(simple,.325)
				(coalA,.075)
				(coalAB,.081)
				(coalABT,.055)
				(alignABT,.055)
			};	
			\end{axis}
			\end{tikzpicture} 
		}
		\subfloat{
			\begin{tikzpicture}
			\begin{axis}[symbolic x coords={simple,coalA,coalAB,coalABT,alignABT},
			xtick={simple,coalA,coalAB,coalABT,alignABT},
			xticklabel style={text height=2ex},
			ymin=0,
			ylabel={Execution time (ms)}, xlabel={Kernel},
			ylabel style={at={(axis description cs:0.1,.5)},anchor=south},
			legend entries={$N=1408$,$N=2048$},
			legend style={at={(0.6,0.45)},anchor=west},
			scaled ticks=false
			]
			
			\addplot [mark=*, color=brown] coordinates {
				(simple,46.920)
				(coalA,8.783)
				(coalAB,9.238)
				(coalABT,5.307)
				(alignABT,5.272)
			};
			\addplot [mark=star] coordinates {
				(simple,130.682)
				(coalA,28.602)
				(coalAB,27.939)
				(coalABT,16.179)
				(alignABT,16.071)
			};	
			\end{axis}
			\end{tikzpicture} 
		}
		\caption{Execution time of matrix multiplication kernels (NVIDIA Tesla P100)}
		\label{fig: matmul runtime}
	\end{figure*}
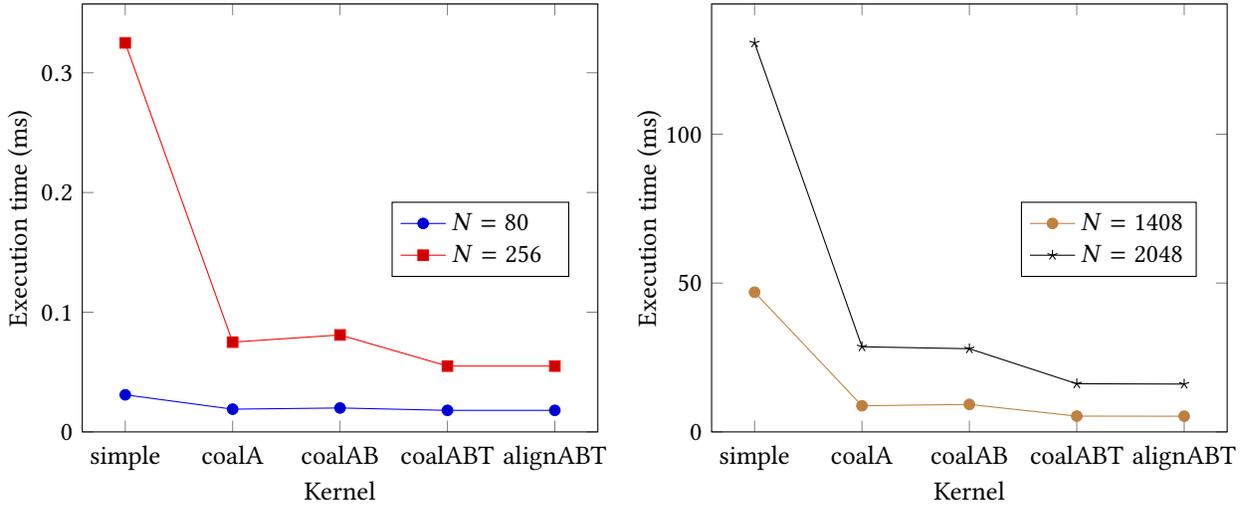
	
	\section{Non-Parallel Memory Metrics} \label{method}
	
	We first consider AIWC memory metrics that do not take into account the interaction between work items executing in parallel.
	We compare these metrics for the incrementally optimized versions of the matrix multiplication example presented in section~\ref{motivating example}.
	
	In addition to the memory metrics reported in \cite{beauaiwc}, we added a new metric of relative local memory usage.
	This measures the proportion of all memory accesses from the symbolic execution of the kernel that occurred to memory allocated as \texttt{\_\_local}.
	On NVIDIA GPUs, this memory address space is mapped to fast on-chip shared memory.
	Relative local memory usage is an example of a metric that is useful to measure performance-critical access patterns on some architectures such as GPUs, and not others, such as CPUs.
	CPUs do not typically have a notion of user-controlled on-chip memory shared between hardware threads such as NVIDIA GPUs' shared memory.
	This is a natural consequence of programming for a heterogeneous system. Specific code patterns may translate to performance improvements only on certain hardware.
	
	In the original AIWC tool \cite{beau_johnston_2017_1134175}, global and local memory address entropy (MAE) were calculated using physical addresses of memory used by the Oclgrind simulator back-end of AIWC. This caused irregularities in entropy calculations across multiple runs of the same simulation. We improved the calculation of memory address entropy by using \textit{virtual addresses} to calculate MAE values using an abstract ideal address space on which all memory accesses by the kernel occur. This allows AIWC to accurately abstract over the hardware and ISA-specific differences in memory layouts across the diverse hardware targets.

	Figure \ref{fig: matmul runtime} shows execution times for the matrix multiplication kernels presented in section~\ref{motivating example} for $N \times N$ matrices of different sizes.
	We recorded kernel execution time exclusive of data transfer on an NVIDIA Tesla P100 using NVIDIA OpenCL 1.2 CUDA 10.1.236 (driver 418.87), and took the average of 100 runs.
	Across varying problem sizes, the vendor recommended optimizations to the matrix multiplication code lead to increased performance. 

	\begin{table*}[h!]
	\centering
	\begin{tabular}[t]{lrrrrr}
		\toprule
		& simple & coalescedA & coalescedAB & coalescedABT & alignedABT \\ \midrule 
		Total memory footprint        & 196608 & 196608     & 196608      & 196608       & 196608     \\ 
		90\% Memory Footprint         & 118196 & 56176      & 489         & 489          & 489        \\ 
		Global MAE                    & 17.02  & 13.18      & 9.78        & 9.78         & 9.78       \\ 
		LMAE \#bits=3                 & 16.02  & 12.18      & 8.78        & 8.78         & 8.78       \\ 
		LMAE \#bits=10                & 9.02   & 5.18       & 1.78        & 1.78         & 1.78       \\ 
		Relative Local Memory Usage   & 0      & 0.50       & 0.94       &  0.94         & 0.94      \\ \bottomrule
	\end{tabular}
	\caption{Selection of AIWC\cite{beauaiwc} metrics for 256 $\times$ 256 matrix multiplication}
	\label{matmul aiwc table}
	\end{table*}

	Table \ref{matmul aiwc table} summarizes AIWC memory metrics collected for each of the matrix multiply kernels in section \ref{motivating example}. Note that only two LMAE values are shown for brevity. We now analyse the effectiveness of AIWC metrics in profiling the NVIDIA recommended memory optimizations applied to the matrix multiplication kernel.
	
	\textit{Relative local memory usage}: As reliance on NVIDIA GPUs' shared memory increases in each kernel from \texttt{simple} to \texttt{coalescedAB}, we find that the proportion of OpenCL local memory usage increases as expected.
	
	\textit{Global and local MAE}: Entropy measurements decrease from \texttt{simple} to \texttt{coalescedAB}, as the optimizations reduce the number of reads from matrices $A$ and $B$ in global memory, replacing these with reads from smaller tiles in local memory.
	We observe an almost completely uniform distribution of memory accesses in \texttt{simple}, where the program makes $N$ loads to each element of $A$ and $B$ with $N$ the dimension of the matrices.
	The distribution of memory accesses becomes increasingly non-uniform as we perform fewer accesses to matrices $A$ and $B$ and more accesses to smaller local memory buffers, resulting in decreases in local and global memory entropy values from \texttt{simple} to \texttt{coalescedAB}.
	
	\textit{Memory Footprint}: Similar to trends in global and local MAE, we find that the ratio of 90\% memory footprint to total memory footprint decreases from $60.12\%$ for \texttt{simple} to $0.25\%$ (\texttt{coalescedAB}). Increased utilization of local memory in the optimized kernels means that the local memory buffers make up a greater proportion of total memory accesses in the program. As the local memory buffers are small and reused within a workgroup, the memory footprint of the local memory buffers is also smaller. 
	
	AIWC's metrics strongly reward optimizations that tend to localize memory accesses. Local memory buffers are typically smaller than global memory arrays when programming for GPUs due to hardware limitations on sizes of shared memory \cite{cudamanual}. However, the metrics currently measured by AIWC do not have a direct causal relation to code patterns that optimize memory accesses on GPUs. The proposed relative local memory usage metric is the first to correspond to a recommended optimization strategy of using fast on-chip shared memory. Further, we find that all current AIWC metrics do not measure any sizeable difference between \texttt{coalescedAB}, \texttt{coalescedABT} and \texttt{alignedABT} codes.
	We address this by proposing another new metric for locality of memory accesses in the following section.
	
	\section{A Parallel Spatial Locality Metric} \label{discussion}
	
	Aggregate metrics of the kind presented by AIWC necessarily present a simplified view of program behaviour, omitting many details.
	Different ways of aggregating program measurements lead to different features of program execution being emphasised in the final metrics.
	For example, the calculation of memory address entropy described in section~\ref{related work} relies only on the frequency distribution of memory accesses to all addresses accessed by the kernel, and discards temporal information.
	The order of sequential memory accesses performed by each thread, as well the relationship between work items in an OpenCL work group, are both vital in accurately characterizing parallel codes.

	We propose a new architecture-independent metric, \emph{parallel spatial locality}, to measure memory access patterns in parallel programs.
	The proposed metric is inspired by CuMAPz' direct approaches to measuring optimization specific characteristics of CUDA codes.
	
	During simulation, AIWC collects a list of all memory accesses by each thread of execution.
	In the OpenCL programming model, threads within a work groups execute the same code and share access to \textit{local memory}.
	We can group together memory accesses of threads in a work group at each logical time step in the symbolic execution of the code.
	On a GPU, memory accesses executed at the same time by different threads in a work group are likely to interact, determining the extent of memory access coalescing and bank conflicts.
	
		\begin{figure*}[h!]
		\centering
		\begin{tikzpicture}
		\begin{axis}[symbolic x coords={0,1,2,3,4,5,6,7,8,9,10},
		width=0.7\textwidth,
		height=0.5\textwidth,
		xtick={0,1,2,3,4,5,6,7,8,9,10},
		xticklabel style={text height=2ex},
		ymin=0,
		ylabel={$n$-bits dropped Parallel Spatial Locality}, xlabel={\# of bits skipped},
		ylabel style={at={(axis description cs:0.05,.5)},anchor=south},
		legend entries={\texttt{simple},\texttt{coalescedA},
			\texttt{coalescedAB}, \texttt{coalescedABT}, \texttt{alignedABT}},
		legend style={at={(0.55,0.75)},anchor=west}
		]
		
		\addplot coordinates {
			(0, 3.987)
			(1, 3.987)
			(2, 3.987)
			(3, 3.49)
			(4, 2.992)
			(5, 2.494)
			(6, 1.996)
			(7, 1.996)
			(8, 1.996)
			(9, 1.996)
			(10, 1.996)
		};

		\addplot coordinates {
			(0,3.952)
			(1,3.952)
			(2,3.952)
			(3,3.459)
			(4,2.965)
			(5,2.472)
			(6,1.979)
			(7,1.506)
			(8,1.033)
			(9,0.5598)
			(10,0.08691)
		};
		
		\addplot coordinates {
			(0,4.145)
			(1,4.145)
			(2,4.145)
			(3,3.628)
			(4,3.11)
			(5,2.598)
			(6,2.075)
			(7,1.603)
			(8,1.13)
			(9,0.6579)
			(10,0.1854)
		};
		
		\addplot [mark=*, color=red] coordinates {
			(0,4.145)
			(1,4.145)
			(2,4.145)
			(3,3.2)
			(4,2.254)
			(5,1.308)
			(6,0.3621)
			(7,0.318)
			(8,0.2738)
			(9,0.2296)
			(10,0.1854)
		};
		
		\addplot coordinates {
			(0,4.145)
			(1,4.145)
			(2,4.145)
			(3,3.2)
			(4,2.254)
			(5,1.308)
			(6,0.3621)
			(7,0.318)
			(8,0.2738)
			(9,0.2296)	
			(10,0.1854)
		};
		\end{axis}
		\end{tikzpicture}
		\caption{Parallel spatial locality metric obtained from AIWC for matrix multiply kernels for 256 $\times$ 256 matrix multiplication}
		\label{fig: matmul parallel spatial locality}
	\end{figure*}
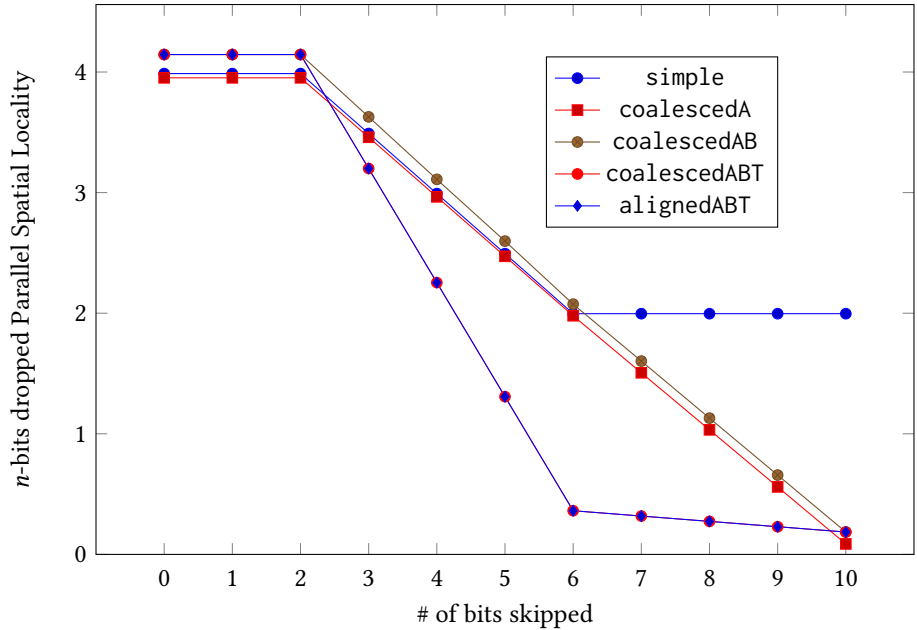
	
	There are three steps involved in generating an AIWC metric: \textit{recording}, \textit{calculating} and \textit{summarizing} data collected from the symbolic execution of the kernel under inspection.
	
	\textit{Record}: we first record memory accesses performed by each thread in an OpenCL work group as described above to achieve a global ordering of all memory accesses performed by the group.
	This ordering is collected in the form of logical timestamps ($t_0 .. t_{last}$) at which memory accesses occur.
	
	\textit{Calculate}: for each timestamp $t = t_0 .. t_{last}$, calculate the $n$-bits-dropped entropy of memory addresses accessed by all threads in a work group within the timestamp $t$. Here $n$ can range between 0 and 10 as was the case for $LMAE$.
	
	\textit{Summarize}: average the collected entropy values across all the timestamps to calculate the parallel spatial locality metrics for one thread group. We then average the $n$-bits-dropped entropy summaries across thread groups to obtain the $n$-bits-dropped parallel spatial locality metrics for the kernel's execution.
	
	

	The proposed metric is a parallel computing analogue for MICA's data stride metric that measures the distance between consecutive data accesses in a single-threaded environment. 
	In parallel programs, to accurately measure spatial locality of accesses, we must consider memory accesses performed by multiple threads in a close temporal scope. 
	The proposed metric calculates the locality of accesses in each time step of the program's execution and steeper reductions in $n$-bits-dropped parallel spatial locality scores will be observed in programs that often access nearby memory addresses within the same timestamp.
	Such programs will perform better on GPUs, as they will make better use of both global memory access coalescing and shared memory bank structures.
	To a lesser extent, the proposed metric reflects performance-critical memory access patterns on CPUs, as pulling a single cache line from global memory into last-level cache may improve memory access times for all CPU cores.
	
	Figure~\ref{fig: matmul parallel spatial locality} shows the proposed parallel spatial locality metric as measured by AIWC for each of the matrix multiplication kernels from section~\ref{motivating example}.
	We observe that the \texttt{coalescedABT} and \texttt{alignedABT} kernels have the steepest reductions in entropy as the number of bits skipped is increased, which correlates to better locality of parallel memory accesses.
	It is expected for these kernels to exhibit better parallel spatial locality than \texttt{simple}, as \texttt{coalescedABT} and \texttt{alignedABT} make use of local memory, where accesses tend to be localized simply due to the small size of shared on-chip memory typically available on GPUs.
	Further we find that the proposed metric successfully distinguishes between the \texttt{coalescedAB} and \texttt{coalescedABT} kernels.
	It accurately depicts a steeper reduction for the more optimized \texttt{coalescedABT} kernel, where a larger proportion of parallel memory accesses make better use of the memory bank structure of GPU shared memory than all previous kernels.
	This is a significant improvement over the state-of-the-art AIWC metrics in characterizing how codes localize simultaneous memory accesses to better use the hardware provided.
	
	\section{Evaluation: Extended OpenDwarfs Benchmark Suite} \label{results}
	
	The Extended OpenDwarfs benchmark suite~\cite{johnston18opendwarfs, krommydas2016opendwarfs} is a set of diverse OpenCL workloads.
	Each benchmark is assigned to one of the 13 Berkeley Dwarfs, common computational and communication patterns which aim to capture the landscape of parallel computing workloads \cite{asanovic2006landscape}. 
	To show that the proposed parallel spatial locality metric is useful for understanding performance properties of a wide range of application codes, we present the results of running the AIWC tool on selected benchmarks from the suite~\cite{opendwarfs2020head}.
	The workloads presented are not optimized for any specific architecture -- hence optimizations using OpenCL local memory (which translates to CUDA shared memory) are not performed. 
	
	Many of these benchmarks can be run with up to four problem sizes based on the sizes of caches found in modern CPU memory hierarchies~\cite{johnston18opendwarfs}.
	For these results we used a problem size setting of \emph{medium}, except for the GEM benchmark, which is run at a setting of \emph{tiny}.
	For benchmarks such as BFS that have multiple kernel invocations per run, we present the AIWC parallel spatial locality metrics for the invocation with the highest number of LLVM-SPIR instructions executed.
	Note that the presented benchmarks use 32-bit numeric types (OpenCL \texttt{int} and \texttt{float}), so dropping up to 2 bits of the memory addresses accessed will not change the parallel spatial locality, since any addresses accessed are at least 4 bytes apart.
	
	
	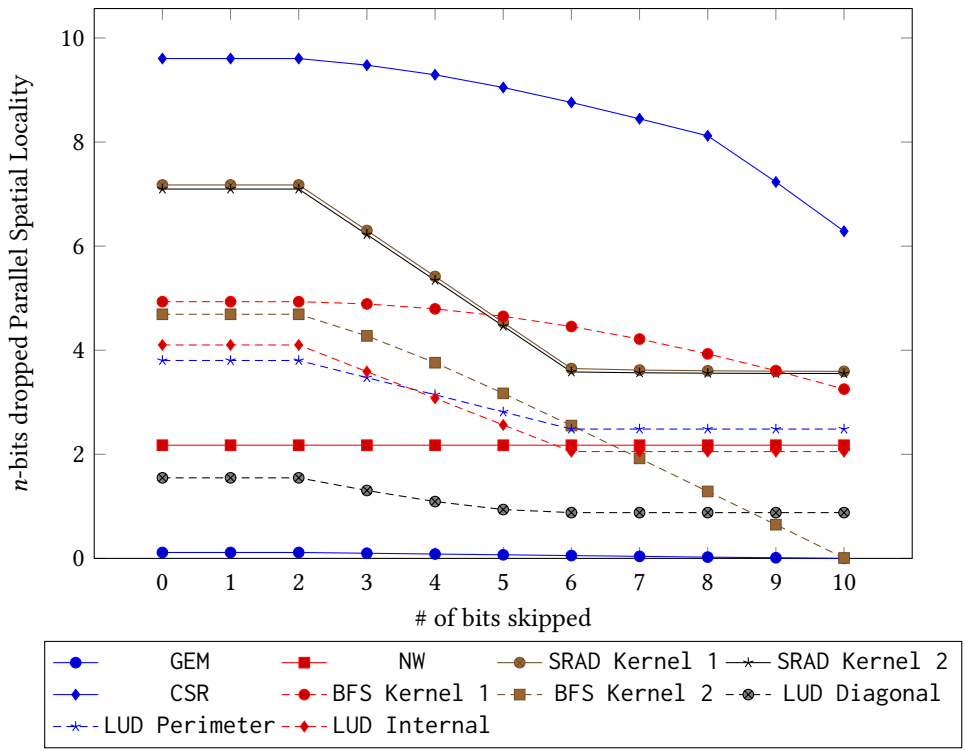
\begin{figure*}[h!]
		\centering
		\begin{tikzpicture}
		\begin{axis}[symbolic x coords={0,1,2,3,4,5,6,7,8,9,10},
		width=0.7\textwidth,
		height=0.5\textwidth,
		xtick={0,1,2,3,4,5,6,7,8,9,10},
		xticklabel style={text height=2ex},
		ymin=0,
		ylabel={$n$-bits dropped Parallel Spatial Locality}, xlabel={\# of bits skipped},
		legend columns=4,
		ylabel style={at={(axis description cs:0.05,.5)},anchor=south},
		legend entries={\texttt{GEM}, \texttt{NW}, \texttt{SRAD Kernel 1},\texttt{SRAD Kernel 2},
			\texttt{CSR}, \texttt{BFS Kernel 1}, \texttt{BFS Kernel 2}, \texttt{LUD Diagonal}, \texttt{LUD Perimeter}, \texttt{LUD Internal}},
		legend style={at={(0.5,-0.15)},anchor=north}
		]

		\addplot coordinates {
			(0, 0.1139)
			(1, 0.1139)
			(2, 0.1139)
			(3, 0.0989)
			(4, 0.08402)
			(5, 0.06914)
			(6, 0.05431)
			(7, 0.03938)
			(8, 0.02442)
			(9, 0.0124)
			(10, 0.00038)
		};

		\addplot coordinates {
			(0, 2.175)
			(1, 2.175)
			(2, 2.175)
			(3, 2.175)
			(4, 2.175)
			(5, 2.175)
			(6, 2.175)
			(7, 2.175)
			(8, 2.175)
			(9, 2.175)
			(10, 2.175)
		};
		
		\addplot coordinates {
			(0, 7.176)
			(1, 7.176)
			(2, 7.176)
			(3, 6.301)
			(4, 5.418)
			(5, 4.533)
			(6, 3.648)
			(7, 3.619)
			(8, 3.605)
			(9, 3.598)
			(10, 3.594)
		};

		\addplot coordinates {
			(0, 7.097)
			(1, 7.097)
			(2, 7.097)
			(3, 6.223)
			(4, 5.344)
			(5, 4.464)
			(6, 3.583)
			(7, 3.567)
			(8, 3.559)
			(9, 3.554)
			(10, 3.553)
		};
		
		\addplot coordinates {
			(0, 9.605)
			(1, 9.605)
			(2, 9.605)
			(3, 9.478)
			(4, 9.293)
			(5, 9.049)
			(6, 8.761)
			(7, 8.447)
			(8, 8.12)
			(9,7.233)
			(10,6.287)
		};

		\addplot coordinates { 
			(0,4.933)	
			(1,4.933)
			(2,4.933)
			(3,4.89)
			(4,4.795)
			(5,4.652)
			(6,4.456)
			(7,4.215)
			(8,3.93)
			(9,3.608)
			(10,3.253)
		};		
	
		\addplot coordinates { 
			(0,4.691)
			(1,4.691)
			(2,4.691)
			(3,4.276)
			(4,3.760)
			(5,3.171)
			(6,2.552)
			(7,1.922)
			(8,1.286)
			(9,0.648)
			(10,0.00776)
		};
		
		\addplot coordinates { 
			(0,1.548)
			(1,1.548)
			(2,1.548)
			(3,1.305)
			(4,1.092)
			(5,0.9385)
			(6,0.8796)
			(7,0.8796)
			(8,0.8796)
			(9,0.8796)
			(10,0.8796)
		};
	
	    \addplot coordinates { 
	    	(0,3.803)
	    	(1,3.803)
	    	(2,3.803)
	    	(3,3.473)
	    	(4,3.144)
	    	(5,2.814)
	    	(6,2.485)
	    	(7,2.485)
	    	(8,2.485)
	    	(9,2.485)
	    	(10,2.485)
    	};
   
   		\addplot coordinates { 
   			(0,4.101)
   			(1,4.101)
   			(2,4.101)
   			(3,3.589)
   			(4,3.077)
   			(5,2.565)
   			(6,2.053)
   			(7,2.053)
   			(8,2.053)
   			(9,2.053)
   			(10,2.053)
   		};
	
		\end{axis}
		\end{tikzpicture}
		\caption{Parallel spatial locality metric for selected OpenDwarfs benchmark kernels}
		\label{fig: opendwarfs parallel spatial locality}
	\end{figure*}

	\subsection{N-body methods: GEM}
	
	The GEM benchmark computes the electrostatic potential of a biomolecule by calculating the sum of charges contributed by all atoms in the biomolecule at each specific surface vertex.
	This is an embarrassingly parallel problem. Each OpenCL work-item operates on a single surface vertex, finding the electrostatic potential generated by looping through every atom in the biomolecule in global order. 
	The computation pattern is highly regular and memory accesses are perfectly synchronized. Atom data is accessed consecutively, with all work-items simultaneously accessing each atom's data.
	The parallel spatial locality metric reflects this pattern of efficient memory utilization (single loads from global memory servicing all OpenCL work-items). The recorded parallel spatial locality approaches the theoretical limit of 0 (0.0124 at 10 bits dropped). This indicates that almost all memory accesses made by the kernel are perfectly synchronized between OpenCL threads.
	Performance results \cite{johnston18opendwarfs, krommydas2016opendwarfs} show that GEM performs significantly better on GPUs than on CPUs, as memory unit stalls are at low levels for both CPUs and GPUs due to the highly efficient memory utilization of this benchmark. 
	As memory operations do not present a bottleneck, this benchmark is able to take advantage of the superior floating-point compute capability of GPUs~\cite{krommydas2016opendwarfs}. 
	
	\subsection{Dense Linear Algebra: Lower-upper decomposition (LUD)}
	
	LUD in OpenDwarfs is a program to decompose an input $N\times N$ matrix as the product of one lower and one upper diagonal matrix. 
	Memory access patterns in a dense linear algebra workload such as LUD are typically highly regular and deterministic for each OpenCL work item, based on the matrix dimension and offset parameters. The OpenDwarfs implementation of LUD \cite{opendwarfs2020head} splits the LUD computation into three kernels: \texttt{LUD perimeter} and \texttt{LUD diagonal} kernels spawn work-items in a single work dimension; while \texttt{LUD internal} is decomposed in both dimensions of the input.
	
	The LUD kernels partially benefit from memory access coalescing on GPUs, for the lines of code where all the work-items access contiguous memory along a row of the input matrix. 
	However, when threads simultaneously access a column of the input matrix, multiple memory requests are made as addresses accessed are too distant to be coalesced into a single memory transaction on GPUs. 
	This is reflected in the drop in parallel spatial locality for the LUD kernels to approximately half the value at 0 bits skipped. Taking \texttt{LUD internal} as an example, the 0-bits skipped parallel spatial locality metric is 4.101, while the 10-bits skipped parallel spatial locality metric is 2.053. 
	The swift decline of the metric to 2.053 from 2 to 6-bits skipped parallel spatial locality indicates that approximately half the parallel memory accesses in \texttt{LUD internal} are highly localized. This occurs when work-items simultaneously access contiguous memory along a matrix row. 
	Similar to the presented matrix multiply codes, LUD can be optimized for GPUs by effective utilization of shared memory \cite{opendwarfs2020head}.
	
	
	\subsection{Dynamic Programming: Needleman-Wunsch (NW)}
	
	Needleman-Wunsch is a dynamic programming algorithm used to perform protein sequence alignment by identifying the similarity between two strings of amino acids.
	The computation of each element in the similarity matrix depends on its west, north and north-west neighbours.
	This dependency enforces a wavefront computation pattern, travelling along the main diagonal of the matrix. Each iteration of the kernels computes over an antidiagonal of the matrix, starting at its top left corner, and finishing at its bottom right corner.
	
	In particular, the wavefront pattern of computations causes each work-item to request distant memory addresses in a parallel fashion. Thus, parallel memory requests into the input matrix prohibit locality. We note the lack of any dropoff in parallel spatial locality as the number of bits dropped increases. This rightly indicates that memory addresses accessed at each logical timestamp are very distant. On GPUs, this translates to poor utilization of both memory access coalescing and caching \cite{krommydas2016opendwarfs}. This trend in the parallel spatial locality metric suggests that a possible improvement for GPU performance would be to load blocks of the input matrix into on-chip local memory to reduce the number of global memory requests -- this is typically performed in GPU optimized implementations of the Needleman-Wunsch algorithm \cite{opendwarfs2020head}. 
	
	\subsection{Structured Grids: Speckle Reducing Anisotropic Diffusion (SRAD)}
	
	SRAD removes locally correlated noise from images by following a repeated grid update computation pattern on image pixel grids. 
	Conditional statements in the code cause thread divergence, potentially within a work-group, to handle boundary conditions. 
	These boundaries constitute a small portion of the executed work-items, especially on larger data-sets and so the effective thread divergence is minimal.
	Memory access patterns in SRAD are statically determined and relatively localized. 
	Similar to the LUD and \texttt{simple} kernels, both SRAD kernels observe a rapid drop-off in parallel spatial locality (figure \ref{fig: opendwarfs parallel spatial locality}) as the number of bits skipped is increased, with the metric stabilizing at approximately half its original value when 0 bits are skipped.
	
	At each point in the kernels' memory access profile, each work-item with global ID $(i,j)$ in an OpenCL work-group accesses the $(i,j)^{th}$ elements of various matrices. The sequential memory access pattern is non-linear since different matrices are accessed consecutively by the kernel, prohibiting ideal caching \cite{krommydas2016opendwarfs}. However, memory requests made simultaneously by a work-group always fall within a rectangular block of one of the matrices. 
	This allows memory access coalescing on GPUs for OpenCL work-items accessing contiguous matrix data, greatly reducing the number of memory requests made on each line of the kernel code on GPUs. 
	However, work-items accessing data along a column of the matrix do not observe memory access coalescing.
	Thus we observe that while cache hit rates are typically low on this benchmark, particularly on GPUs \cite{krommydas2016opendwarfs}, GPUs can hide the latency of global memory accesses through memory access coalescing to some extent.
	
	\subsection{Graph Traversal: BFS}
	
	Graph traversal algorithms require pointer chasing operations to traverse nodes of a graph and perform calculations. 
	The Breadth-First Search implemented in OpenDwarfs calls two OpenCL kernels to traverse nodes immediately connected to the list of nodes at each depth starting from the root node. 
	BFS is characterized by an imbalanced workload per kernel launch depending on the degree of the nodes being operated on, with only a proportion of launched nodes performing meaningful work. 
	As such, the AIWC features collected for each kernel invocation vary significantly. 
	The parallel spatial locality metric collects the entropy of memory accesses at each time step, and workload imbalances across work-items are dealt with by averaging entropy scores across all the time steps in the execution of an OpenCL work-group. 
	
	Of the two BFS kernels, it is \texttt{BFS kernel 1} that performs the graph traversal necessary to generate a list of neighbours at each node. The memory access patterns in \texttt{BFS kernel 1} are irregular. 
	Work-items fetch discontinuous memory locations attributed to any particular node in the graph, depending on the connectivity of the node they operate on. 
	The program structure involves multiple levels of pointer-chasing and thus the precise memory addresses accessed are data-dependent. 
	This leads to poor parallel spatial locality. Thus, we see a slow dropoff in parallel spatial locality for \texttt{BFS kernel 1}.
	
	\subsection{Sparse Linear Algebra: CSR}
	
	Compressed Sparse Row Matrix-Vector Multiplication (CSR) computes the product of a sparse matrix and a dense vector.
	The matrix is stored in a compressed row storage sparse matrix format, which is very efficient for storage when the number of zero elements is far greater than the number of non-zero elements. Three inputs are provided to the CSR kernel. The non-zero elements of a matrix are stored in row-major order in $Ax$, along with separate arrays $Aj,\ Ap$ indicating the position of each non-zero element in the matrix.
	
	In the OpenDwarfs implementation, each row of the input sparse matrix is assigned to a separate OpenCL work item. 
	The locations of matrix data read by each work-item are dependent on the number and position of non-zero values in the sparse-matrix, which are decided by the values in $Aj$ and $Ap$. 
	Similar to BFS, this means memory access patterns are runtime-dependent due to indirect addressing.
	This pattern of indirect addressing is typical of applications in the Spare Linear Algebra Dwarf, which severely hinders locality of memory accesses performed. The collected parallel spatial locality metric reflects this trend. Figure \ref{fig: opendwarfs parallel spatial locality} shows the gradual decline in parallel spatial locality as the number of bits dropped is increased.
	
	\section{Conclusions and Future Work} \label{future work}
	
	To the best of our knowledge, this work is the first to propose the use of architecture-independent metrics of parallel memory access to guide hardware-specific optimization. 
	We implemented and evaluated a new feature, relative local memory usage, to help characterize memory-based GPU code optimizations.
	We also implemented a new parallel spatial locality metric to capture the idea of closeness of memory accesses made by parallel OpenCL workloads.
	We ran the enhanced AIWC tool on matrix multiply kernels and selected OpenDwarfs benchmarks, presenting results and analysis to validate our methodology. 
	
	Our proposed parallel spatial locality metric may also correlate to some memory-based optimization strategies for CPUs. 
	Future work would apply the approach followed by this paper to optimizations for CPU and FPGA architectures to critically evaluate the viability of AIWC in analyzing memory access patterns in codes targeted to these architectures.
	
	While our work in this paper intended to guide optimization efforts, another potential use case for AIWC is providing architecture-independent performance predictions for OpenCL kernels by generating machine learning models based on AIWC metrics \cite{beauperformanceprediction}.  Future work would modify the presented metrics and develop new memory metrics with the specific intent of being fed to machine learning models to predict kernel performance on arbitrary and novel architectures. 
	
	\bibliographystyle{plain}
	\bibliography{bib/bibliography.bib}

\begin{thebibliography}{10}

\bibitem{opendwarfs2020head}
{Extended OpenDwarfs}.
\newblock
  \url{https://github.com/ANU-HPC/OpenDwarfs/commit/dee488cac9833f029dfada35ff6ae4077b68c4b5},
  Jan 2020.

\bibitem{asanovic2006landscape}
Krste Asanovic, Ras Bodik, Bryan~Christopher Catanzaro, Joseph~James Gebis,
  Parry Husbands, Kurt Keutzer, David~A Patterson, William~Lester Plishker,
  John Shalf, Samuel~Webb Williams, et~al.
\newblock The landscape of parallel computing research: A view from berkeley.
\newblock Technical report, Technical Report UCB/EECS-2006-183, EECS
  Department, University of California, Berkeley, 2006.

\bibitem{ganesan2008performance}
Karthik Ganesan, Lizy John, Valentina Salapura, and James Sexton.
\newblock A performance counter based workload characterization on {Blue
  Gene/P}.
\newblock In {\em International Conference on Parallel Processing ({ICPP})},
  pages 330--337. IEEE, 2008.

\bibitem{hennessycomparch}
John~L. Hennessy and David~A. Patterson.
\newblock {\em {Computer Architecture - A Quantitative Approach, 5th Edition}}.
\newblock Morgan Kaufmann, 2012.

\bibitem{hoste2007microarchitecture}
Kenneth Hoste and Lieven Eeckhout.
\newblock Microarchitecture-independent workload characterization.
\newblock {\em IEEE Micro}, 27(3), 2007.

\bibitem{inteloptimisation}
{Intel Corporation}.
\newblock {Intel® 64 and IA-32 Architectures Optimization Reference Manual}.
\newblock 2016.

\bibitem{beauperformanceprediction}
Beau Johnston, Gregory Falzon, and Josh Milthorpe.
\newblock Opencl performance prediction using architecture-independent
  features.
\newblock In {\em 2018 International Conference on High Performance Computing
  {\&} Simulation, {HPCS} 2018, Orleans, France, July 16-20, 2018}, pages
  561--569. {IEEE}, 2018.

\bibitem{beauaiwc}
Beau Johnston and Josh Milthorpe.
\newblock {AIWC:} {OpenCL}-based architecture-independent workload
  characterization.
\newblock {\em IEEE/ACM Workshop on the LLVM Compiler Infrastructure in HPC
  (LLVM-HPC)}, Nov 2018.

\bibitem{johnston18opendwarfs}
Beau Johnston and Josh Milthorpe.
\newblock {Dwarfs} on accelerators: Enhancing {OpenCL} benchmarking for
  heterogeneous computing architectures.
\newblock In {\em Proceedings of the $47^{th}$ International Conference on
  Parallel Processing Companion}, ICPP '18, pages 4:1--4:10, New York, NY, USA,
  2018. ACM.

\bibitem{beau_johnston_2017_1134175}
Beau Johnston, James Price, Moritz Pflanzer, Petros Kalos, Tom Deakin, Nido
  Media, and Daniel Saier.
\newblock {BeauJoh/Oclgrind: Adding AIWC -- An Architecture Independent
  Workload Characterisation Plugin}.
\newblock https://doi.org/10.5281/zenodo.1134175, December 2017.

\bibitem{kim2011cumapz}
Yooseong Kim and Aviral Shrivastava.
\newblock {CuMAPz}: a tool to analyze memory access patterns in {CUDA}.
\newblock In Leon Stok, Nikil~D. Dutt, and Soha Hassoun, editors, {\em
  Proceedings of the 48th Design Automation Conference, {DAC} 2011, San Diego,
  California, USA, June 5-10, 2011}, pages 128--133. {ACM}, 2011.

\bibitem{krommydas2016opendwarfs}
Konstantinos Krommydas, Wu-chun Feng, Christos~D Antonopoulos, and Nikolaos
  Bellas.
\newblock {OpenDwarfs}: Characterization of dwarf-based benchmarks on fixed and
  reconfigurable architectures.
\newblock {\em Journal of Signal Processing Systems}, 85(3):373--392, 2016.

\bibitem{luk2005pin}
Chi-Keung Luk, Robert Cohn, Robert Muth, Harish Patil, Artur Klauser, Geoff
  Lowney, Steven Wallace, Vijay~Janapa Reddi, and Kim Hazelwood.
\newblock Pin: building customized program analysis tools with dynamic
  instrumentation.
\newblock In {\em {ACM SIGPLAN} notices}, volume~40, pages 190--200. ACM, 2005.

\bibitem{cudaoptimisation}
{NVIDIA Corporation}.
\newblock {CUDA C++ Best Practices Guide}.
\newblock 2019.

\bibitem{cudamanual}
{NVIDIA Corporation}.
\newblock {CUDA C++ Programming Guide}.
\newblock 2019.

\bibitem{price:15}
James Price and Simon McIntosh-Smith.
\newblock Oclgrind: An extensible {OpenCL} device simulator.
\newblock In {\em Proceedings of the 3rd International Workshop on {OpenCL}},
  page~12. ACM, 2015.

\bibitem{shao2013isa}
Yakun~Sophia Shao and David Brooks.
\newblock {ISA}-independent workload characterization and its implications for
  specialized architectures.
\newblock In {\em IEEE International Symposium on Performance Analysis of
  Systems and Software {(ISPASS)}}, pages 245--255. IEEE, 2013.

\bibitem{spafford2010maestro}
Kyle Spafford, Jeremy Meredith, and Jeffrey Vetter.
\newblock Maestro: data orchestration and tuning for {OpenCL} devices.
\newblock {\em Euro-Par 2010-Parallel Processing}, pages 275--286, 2010.

\end{thebibliography}
	
	\pagebreak
	
	\begin{appendix}
		
		\section{\texttt{simple} kernel} \label{simpleMultiply}
		
		\begin{lstlisting}
		__kernel void simpleMultiply(__global float *A,
				__global float *B, 
				__global float *C, 
				int N)
		{
			const int globalRow = get_global_id(0); // Row ID of C (0..N-1)
			const int globalCol = get_global_id(1); // Col ID of C (0..N-1)
			// Compute a single element of C (loop over K)
			float acc = 0.0f;
			for (int k = 0; k < N; ++k) {
				acc += B[k * N + globalCol] * A[globalRow * N + k];
		`	}
			// Store the result
			C[globalRow * N + globalCol] = acc;
		}
		
		\end{lstlisting}
		
		\section{\texttt{coalescedA} kernel} \label{coalescedAMultiply}

		\begin{lstlisting}
		__kernel void coalescedAMultiply(
				const __global float* A,
				const __global float* B,
				__global float* C,
				const int N)
		{
			__local float aTile[TILE_DIM][TILE_DIM];
			
			const int localRow = get_local_id(0);
			const int localCol = get_local_id(1);
			
			const int globalRow = get_global_id(0);
			const int globalCol = get_global_id(1);
			
			__private float sum = 0.0f;
			
			const int numTiles = N / TILE_DIM;
			__private const int tiledRow = globalRow*N+ localCol;
			for (int i = 0; i < numTiles; i++) {
				aTile[localRow][localCol] = A[tiledRow+i*TILE_DIM];
				barrier(CLK_LOCAL_MEM_FENCE);
				for (int k = 0; k < TILE_DIM; k++) {
					sum += aTile[localRow][k] * B[(i*TILE_DIM+k)*N+globalCol];
				}
				barrier(CLK_LOCAL_MEM_FENCE);
			}
			C[globalRow*N+globalCol] = sum;
		}
		\end{lstlisting}

		\section{\texttt{coalescedAB} kernel} \label{coalescedABMultiply}
			
		\begin{lstlisting}
		__kernel void coalescedABMultiply(
				const __global float* A,
				const __global float* B,
				__global float* C,
				const int N) {
			__local float ASub[TILE_DIM][TILE_DIM];
			__local float BSub[TILE_DIM][TILE_DIM];    
			
			const int localRow = get_local_id(0);
			const int localCol = get_local_id(1);
			const int globalRow = get_global_id(0);
			const int globalCol = get_global_id(1);
			
			float acc = 0.0f;
			const int numTiles = N/TILE_DIM;
			
			for (int i = 0; i < numTiles; i++) {
				const int tiledRow = globalRow*N+i*TILE_DIM + localCol;
				const int tiledCol = globalCol + (TILE_DIM*i + localRow)*N;
				ASub[localRow][localCol] = A[tiledRow];
				BSub[localRow][localCol] = B[tiledCol];
				
				barrier(CLK_LOCAL_MEM_FENCE);
				
				for (int k=0; k<TILE_DIM; k++) {
					acc += ASub[localRow][k] * BSub[k][localCol];
				}
				
				barrier(CLK_LOCAL_MEM_FENCE);
			}
			C[globalRow*N + globalCol] = acc;
		}
		\end{lstlisting}
	\end{appendix}
	
\end{document}